\documentclass[aps,superscriptaddress,floatfix,onecolumn]{revtex4}
\usepackage{amsfonts,color,bigints}
\usepackage{amsmath}
\usepackage{amssymb}
\usepackage{physics}
\usepackage{graphicx,dsfont}
\usepackage[utf8]{inputenc}
\usepackage{subfigure}

\newcommand{\element}[3]{\langle #1|#2|#3\rangle}
\newcommand{\spin}{\hat{\mathcal{S}}}

\begin{document}

\title{Excited-State Quantum Phase Transitions in the anharmonic Lipkin-Meshkov-Glick Model I: Static Aspects}

\author{J. Gamito}
\affiliation{Departamento de F\'isica At\'omica, Molecular y Nuclear, Facultad de F\'isica, Universidad de Sevilla, Apartado 1065, E-41080 Sevilla, Spain}

\author{J. Khalouf-Rivera}
\affiliation{Departamento de Ciencias Integradas y Centro de Estudios Avanzados en Física, Matemáticas y Computación, Universidad de Huelva, Huelva 21071, Spain}

\author{J.M. Arias}
\affiliation{Departamento de F\'isica At\'omica, Molecular y Nuclear, Facultad de F\'isica, Universidad de Sevilla, Apartado 1065, E-41080 Sevilla, Spain}
\affiliation{Instituto Carlos I de F\'{\i}sica Te\'orica y Computacional, Universidad de Granada, Fuentenueva s/n, 18071 Granada, Spain \\}

\author{P. P\'erez-Fern\'andez}
\affiliation{Instituto Carlos I de F\'{\i}sica Te\'orica y Computacional, Universidad de Granada, Fuentenueva s/n, 18071 Granada, Spain \\}
\affiliation{Dpto. de F\'isica Aplicada III, Escuela T\'ecnica Superior de Ingenier\'ia, Universidad de Sevilla, Sevilla, Spain }

\author{F. Pérez-Bernal}
\affiliation{Departamento de Ciencias Integradas y Centro de Estudios Avanzados en Física, Matemáticas y Computación, Universidad de Huelva, Huelva 21071, Spain}
\affiliation{Instituto Carlos I de F\'{\i}sica Te\'orica y Computacional, Universidad de Granada, Fuentenueva s/n, 18071 Granada, Spain \\}

\begin{abstract}
  The basic Lipkin-Meshkov-Glick model displays a second order ground state quantum phase transition and an excited state quantum phase transition (ESQPT). The inclusion of an anharmonic term in the Hamiltonian implies a second ESQPT of a different nature. We characterize this ESQPT using the mean field limit of the model. The new ESQPT, associated with the changes in the boundary of the finite Hilbert space of the system, can be properly described using the order parameter of the ground state quantum phase transition, the energy gap between adjacent states, the participation ratio, and the quantum fidelity susceptibility.
\end{abstract}

\maketitle

\section{Introduction}

Ground state quantum phase transitions (QPTs), zero temperature transitions that are triggered by quantum fluctuations instead of thermal ones, have been in the limelight in recent years due to their deep implications in the understanding of many-body quantum systems \cite{Carr2010}. In these transitions the ground state of the system undergoes abrupt and qualitative changes when one or several control parameters in the Hamiltonian straddle a critical value. Since the seminal Gilmore and collaborators works \cite{Gilmore1978, Gilmore1979, Feng1981}, there have been numerous works characterizing QPTs in two-level quantum systems of different dimensionality used to model nuclear and molecular systems, as the interacting boson model (IBM) or the vibron model (see Refs.~\cite{Cejnar2009, Casten2009, Cejnar2010} and references therein). A full classification of ground state QPTs in two-level models with different dimensionality can be found in Ref.~\cite{Cejnar2007}.

The study of QPTs has been extended to the realm of excited states, 
introducing the concept of excited-state quantum phase transition (ESQPT) that implies a non-analyticity in the density of states and the energy level flow which, in most cases, is associated with a ground state QPT and the existence of a critical point in the energy functional obtained from the classical or mean field limit of the system \cite{Cejnar2006, Caprio2008, Cejnar2008}. The nature of the non-analyticity in the energy level density in nondegenerate stationary points of systems with \(n\) effective degrees of freedom, is such that the order of the derivative of the level density that is non-analytic is \(n-1\) \cite{Cejnar2008, Stransky2014, Stransky2015, Stransky2016, Macek2019}. ESQPTs have received a great deal of attention in many systems, fostered by detection of their precursors in molecular spectra \cite{Larese2011, Larese2013, KRivera2019, KRivera2020},  superconducting microwave billiards \cite{Dietz2013}, and spinor condensates \cite{Zhao2014}. In the latter case, some promising developments have been recently published  \cite{Polina2021, Cabedo2021}. Ref.~\cite{Cejnar2021} is a review on different aspects of ESQPTs with an extensive reference list.

A system where ESQPTs have been studied is the two-dimensional limit of the vibron model (2DVM), a two-level model built upon the bilinear products of two cartesian and a scalar bosonic operators \cite{Iachello1996, PBernal2008}. This model has been used to reproduce bending spectra of molecules, as its two dynamical symmetries can be associated with the bending degrees of freedom for linear and bent molecular species, as well as the interesting situations that lie in-between these two limiting cases \cite{Iachello2003, PBernal2005}. The basic 2DVM model Hamiltonian, including only a number and a pairing operator, presents a second order ground state QPT and an associated ESQPT that can be linked to the effect over excited energy levels of the barrier to linearity in nonrigid molecular systems \cite{PBernal2008}. In fact, most experimental ESQPT signatures that were identified in the bending degrees of freedom of nonrigid molecules are associated with this ESQPT \cite{Larese2011, Larese2013, KRivera2020}. To obtain results of spectroscopic quality in the modeling of molecular spectra implies the explicit inclusion of anharmonic terms in the 2DVM Hamiltonian. In Ref.~\cite{PBernal2010}, it was shown that the inclusion of anharmonicity in the Hamiltonian results in a second ESQPT in the broken-symmetry phase, that is where the typical spectroscopic signatures of nonrigid molecules are found. More recently, in the pursuit of the description of the transition state in isomerization reactions \cite{KRivera2019}, it has been found that the second ESQPT  can also be present in the symmetric phase for sufficiently large values of the anharmonicity and it is not associated with the ground state QPT but it stems from changes in the phase-space boundary of the 2DVM's finite-dimensional Hilbert space \cite{KRivera2021}.
 
A particular kind of such ESQPTs have been already studied in matter-radiation interaction models with two degrees of freedom (Dicke and Tavis-Cummings models). In this case, the authors have denoted these ESQPTs happening at the boundary of the model space as \textit{static} ESQPTs, due to their lack of effects on the system dynamics. In this way, they can be distinguished from the more usual  ESQPTs that the authors name as \textit{dynamic} ESQPTs \cite{BMagnani2014}.  More information on ESQPTs associated with the finiteness of the system Hilbert space can be found in Refs.~\cite{Macek2019,Cejnar2021}.

Our aim is to extend the results obtained for the 2DVM \cite{PBernal2010, KRivera2021}, to the Lipkin-Meshkov-Glick (LMG) model. This  model was introduced in 1965 as a toy model for interacting fermions in nuclear physics, to help assessing the performance of different approximations used on that time in the study of nuclear structure \cite{Lipkin1965, Meshkov1965, Glick1965}. It can be mapped to a set of spins with a long range all-to-all interaction, and from their initial purpose it has been later used in many different contexts. In particular, it has been extensively used for the study of  QPTs \cite{Botet1983,Dusuel2004,Dusuel2005,Heiss2005,Leyvraz2005,Castanos2005,Castanos2006,Ribeiro2008,Engelhardt2013,Romera2014,Campbell2016,Heyl2018} and ESQPTs \cite{Relano2008, PFernandez2009, PFernandez2011, Yuan2012, Santos2016, Wang2019a, Wang2019b,GRuiz2021}. 
The general LMG model presents first-, second- and third-order ground state QPTs \cite{Romera2014}, something that has attracted the interest of researchers. In a basic formulation, with a Hamiltonian composed of two operators that can be mapped to a number and a pairing operator, the model has a second order ground state QPT and its associated ESQPT. A particular realization of the LMG model has a classical limit energy functional that is equal to classical limit of the IBM and it has been used as a tool to shed light upon QPTs and ESQPTs in this model \cite{Vidal2006, GRamos2016, GRamos2017}. It is also a toy model that can be of application in quantum computing \cite{Larson2010, Cervia2021, Chinni2021}.
The interest on the LMG model has  been further fostered by the achievement of different experimental realizations, with optical cavities \cite{Morrison2008}, Bose-Einstein condensates \cite{Zibold2010}, nuclear magnetic resonance systems \cite{AFerreira2013}, trapped atoms \cite{Jurcevic2014, Jurcevic2017, Muniz2020}, and cold atoms \cite{Makhalov2019}.

The purpose of the present article is to explore how the LMG model is modified including in the model Hamiltonian the anharmonic  two-body term $\hat n(\hat n + 1)$ with
a negative control parameter $\alpha$. We perform a mean field analysis of the anharmonic LMG model, characterize ground state QPT in this case, and perform a study of the two ESQPTs that appear in the model making use of different quantities such as the energy gap between adjacent levels, the ground state QPT order parameter, the participation ratio, and the quantum fidelity susceptibility. This paper is followed by another one with a focus on the effect of the two ESQPTs on dynamical properties of the model \cite{Gamito2022}.

The present paper is organized as follows. In Sec.~\ref{secII} we introduce the anharmonic LMG model, review its algebraic structure, provide its Hamiltonian matrix elements, and study the model mean field or thermodynamic limit using the coherent state formalism. Sec.~\ref{secIII} includes a description of the different quantities used to characterize the anharmonic LMG ESQPTs and the results obtained for different anharmonicity and system size values. We end with some concluding remarks in Sec.~\ref{secIV}.

\section{The model}
\label{secII}
The LMG model has recently attracted an increasing attention after the experimental realization of one dimensional spin-$\frac{1}{2}$ lattices with a variable interaction range \cite{Richerme2014, Jurcevic2014}. Using Pauli spin matrices, $\sigma_{i,\beta}$ with $i = 1, 2, \ldots, N$, $\beta = x,y,z$, the system Hamiltonian is
\begin{equation}
\hat H^{(a)}= B \sum_{i=1}^{N} \sigma_{i,z} + \sum_{i<j=1}^{N} \frac{K}{|i-j|^a}\sigma_{i,x}\sigma_{j,x}~,
\label{eq:LMG0}
\end{equation}
\noindent where we assume $\hbar = 1$, $B$ is the amplitude of an external magnetic field, $K$ is the energy scale of the interactions between different spin sites, and the $a$ parameter controls the interaction range. For $a = 0$ the interaction range is infinite and Hamiltonian \eqref{eq:LMG0} is mapped to the  Lipkin-Meshkov-Glick (LMG) model \cite{Lipkin1965, Meshkov1965, Glick1965}
\begin{equation}
\hat H=\hat H^{(a = 0)}= (1-\xi)\left(S+\spin_{z}\right)+\frac{2\xi}{S}\left(S^2-\spin_{x}^2\right)~,
\label{eq:LMG1}
\end{equation}
\noindent where we introduce collective spin operators $\spin_{\beta} = \frac{1}{2} \sum_{i=1}^{N}\sigma_{i,\beta}$ for $\beta = x,y,z$, add a constant term $2 B S - K S(2S-1)$, and define a single control parameter, $\xi$, making $2B = 1-\xi$ and $S K = -\xi$. The control parameter is defined in the range $\xi \in [0,1]$, driving the system from one phase to the other one. Indeed, from an algebraic point of view, the LMG Hamiltonian given by Eq.~\eqref{eq:LMG1} presents a $u(2)$ algebraic structure, with two limiting dynamical symmetries: $u(2)\supset u(1)$ and $u(2)\supset so(2)$. Each dynamical symmetry is associated with a different phase. For $\xi=1$, the Hamiltonian is diagonal in the basis associated with the $so(2)$ subalgebra and this is known as the deformed or broken-symmetry phase; whereas for $\xi=0$ the Hamiltonian is diagonal in the  $u(1)$ subalgebra basis and it is in the normal or symmetric phase \cite{frank}. The LMG model Hamiltonian \eqref{eq:LMG1} experiences a second order ground state QPT at the critical value of the control parameter $\xi_c = 0.2$ \cite{Romera2014}.

Inspired by the works Ref.~\cite{PBernal2010, KRivera2021}, we include in the Hamiltonian Eq. \eqref{eq:LMG1} an anharmonic term, using a second-order operator on $\spin_{z}$,

\begin{equation}
\hat H_{anh}=(1-\xi)\left(S+\spin_{z}\right)+\frac{\alpha}{2S}\left(S+\spin_z\right)\left(S+\spin_z+1\right) +\frac{2\xi}{S}\left(S^2-\spin_{x}^2\right)~.
\label{eq:LMG2}
\end{equation}
The new Hamiltonian still depends on the control parameter $\xi$, which drives the system between phases, but it also depends on a second control parameter, $\alpha$. For $\alpha =0$, we recover the original Hamiltonian, Eq.~\eqref{eq:LMG1}, and for $\alpha$ values different from zero, the $\xi=0$ limit is transformed from a truncated one-dimensional harmomic oscillator to an anharmonic oscillator. That is the reason why we denote the 
Hamiltonian \eqref{eq:LMG2} as the anharmonic LMG (aLMG) model. It is worth noticing that the limit $so(2)$ is not recovered any longer for $\xi=1$, unless $\alpha =0$. In Ref.~\cite{Fortunato2010}, a preliminary study of an anharmonic LMG model using only operators diagonal in the $u(1)$ basis was carried out.

The Hilbert space for Hamiltonian Eq.~\eqref{eq:LMG0} has dimension $2^N$, but in the long range interaction \eqref{eq:LMG2} there is a drastic reduction in the dimension of the Hilbert space  due to the conservation of the total spin, $[\spin^2,\hat{H}]=0$. We focus on the sector that corresponds to the maximum angular momentum, $S=N/2$, with a Hilbert space dimension $N+1$.

In the aLMG model, as in the original LMG model, there are two bases available to carry out the calculations, one defined by the $u(2)\supset u(1)$ dynamical algebra, $|S,M_z\rangle$ where $M_z=-S,\ldots,0,\ldots,S$ is the projection of the total spin $S$ in the $z$ direction. The matrix elements of the Hamiltonian Eq. \eqref{eq:LMG2} in the $u(1)$ basis are

\begin{eqnarray}
\element{S,M_{z}'}{\hat H_{anh}}{S,M_{z}}&=&\left\{(1-\xi) (S + M_z) + \frac{\xi}{2S}\left[4S^2 - (S-M_z)(S+M_z+1) - (S+M_z)(S-M_z+1)\right]\right. \nonumber \\
& & \left. + \frac{\alpha}{2} \left[S + 1 + \left(2 + \frac{1}{S}\right)M_z + \frac{M_z^2}{s}\right]\right\}\delta_{M_{z}',M_{z}}\label{u1matel} \\
&-&\frac{\xi}{2S}\sqrt{(S-M_z)(S-M_z-1)(S + M_z+2)(S + M_z+1)}\;\delta_{M_{z}',M_{z}+2}~. \nonumber
\end{eqnarray}

As can be easily seen from the matrix elements in Eq.~\eqref{u1matel}, only states with $M_{z}' = M_z$ or  $M_{z}' = M_z\pm 2$ are connected, hence the aLMG Hamiltonian Eq.~\eqref{eq:LMG2} conserves parity, $(-1)^{S+M_z}$ and the Hamiltonian matrix is split into two blocks, one for positive or even parity and dimension $S+1$, and the other for negative or odd parity with dimension $S$.

The second basis is associated with the $u(2)\supset so(2)$ symmetry, $|S,M_x\rangle$, and in this case it is the projection of the spin in the $x$ direction the second quantum label in the basis. The matrix elements in this case can be deduced from the previous ones once the system is rotated \cite{frank}.

\begin{figure}
\begin{centering}
\includegraphics[scale=0.5]{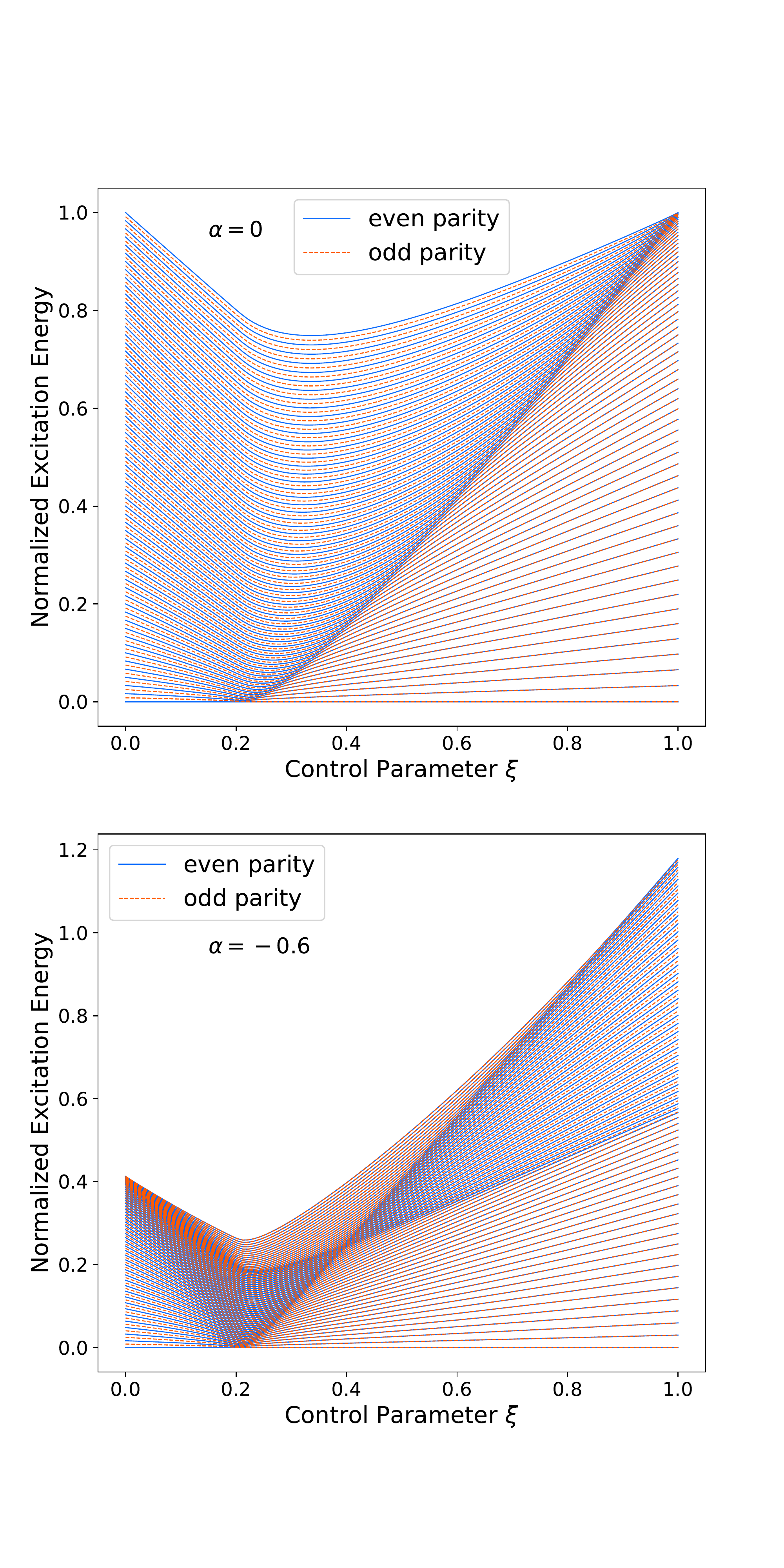}
\par\end{centering}
\caption{(Color online) Correlation energy diagram depicting the normalized excitation energies for even (blue lines) and odd (dashed red lines) parity states as a function of the control parameter $\xi\in[0,1]$ for a system with size $N  = 120$ and anharmonicity parameter $\alpha = 0$ (upper panel) and $-0.6$ (lower panel).}
\label{ced_figure}
\end{figure}

We depict in Fig.~\ref{ced_figure} the correlation energy diagram for the aLMG Hamiltonian Eq.~\eqref{eq:LMG2} with a system size $N = 2S = 120$ and $\alpha = 0$ and $-0.6$ in the upper and lower figure panels, respectively. In both cases, we plot the normalized excitation energy as a function of the control parameter $\xi$. Both positive and negative parity states are included in the figure, positive parity energy levels are plotted with full blue lines and odd parity states with dashed red lines. In the upper panel, it is clearly evinced the ground state QPT at $\xi_c = 0.2$ and the separatrix, marked by a high density of states, that is the boundary between the two ESQPT phases. Excited states above the separatrix have a $u(1)$ (symmetric) character, while those below it have a $so(2)$ (broken symmetry) character. It can be appreciated in the figure how beyond the critical value of the control parameter $\xi$ and for states under the separatrix, the even and odd parity states are degenerated. From this figure it is already clear that the new ESQPT is not a static ESQPT using the notation introduced in Ref.~\cite{BMagnani2014}. In the present case, as it is confirmed in the semiclassical analysis in the next subsection, the non-analyticity of the density of states is of the same kind in the two ESQPTs and both have noticiable effects on the system dynamics \cite{Gamito2022}. 

The lower panel in Fig.~\ref{ced_figure} is the correlation energy diagram for a negative $\alpha$ value. As expected from the results obtained for the 2D limit of the vibron model \cite{PBernal2010, KRivera2021}, there is a new separatrix that crosses the one associated with the ground state QPT and the degeneracy pattern is more complex. In order to better understand the second separatrix we study the classical limit of the model making use of the spin coherent states  formalism \cite{spin_cohsta}.

\subsection{Classical limit of the model}
The classical limit of Hamiltonian Eq.~\eqref{eq:LMG2} can be obtained within a mean-field analysis, studying the large-size limit of the model. We perform this study using spin coherent states \cite{spin_cohsta}. From the usual definitions and using the notation introduced in the previous section,  the spin ladder operators are $\spin_{\pm}=\spin_x\pm i\spin_y$ and the spin coherent state is defined as
\begin{equation}
  \ket{\left[S\right]\mu}=\frac{\exp\left(\mu\spin_+\right)}{\left(1+\left|\mu\right|^2\right)^S} \ket{S,-S}=\frac{1}{\left(1+\left|\mu\right|^2\right)^S} \sum\limits^{2S}_{p=0}\sqrt{\binom{2S}{p}} \mu^p\ket{S,-S+p}  ~, \label{spin_coh_state}
\end{equation}
where $\mu$ is a complex parameter that encompasses the classical position and momentum variables, $\ket{S,-S+p}$ is a eigenstate of $\spin_z$,
$\spin_z\ket{S,-S+p}=(-S+p)\ket{S,-S+p}$, and
$(\spin_+)^p\ket{S,-S}=\sqrt{\frac{p!(2S)!}{(2S-p)!}}\ket{S,-S+p}$
\cite{spin_cohsta}.

The expectation value of the operators $\spin_z$, $\spin_+$, and $\spin_-$ with the spin coherent state \eqref{spin_coh_state} are
\begin{eqnarray}
  \bra{[S]\,\mu} \spin_z \ket{[S]\,\mu} & = & S\frac{|\mu|^2-1}{|\mu|^2+1}~, \nonumber\\
  \bra{[S]\,\mu} \spin_+ \ket{[S]\,\mu} & = & S\frac{2\mu^*}{|\mu|^2+1}~, \\
  \bra{[S]\,\mu} \spin_- \ket{[S]\,\mu} & = & S\frac{2\mu}{|\mu|^2+1} ~.\nonumber
\end{eqnarray}

Thus, the classical limit of aLMG,  $H_{cl}(\mu)$, can obtained as the expectation value of the Hamiltonian Eq.~\eqref{eq:LMG2} (per particle) with the coherent state in the large system size limit ($S \rightarrow \infty$)

\begin{equation} \label{enfun} H_{cl}(\mu) = \frac{\bra{[S]\,\mu}
    \hat{H}_{anh} \ket{[S]\,\mu}}{2S}=\left(1-\xi \right)
  \frac{|\mu|^2}{1+|\mu|^2} + \alpha
  \left(\frac{|\mu|^2}{1+|\mu|^2}\right)^2 +\xi \left[ 1 - \left(
      \frac{\mu+\mu^*}{1+|\mu|^2} \right)^2 \right]\; ,
\end{equation}
\noindent where the complex variable $\mu$ can be mapped into $(q,p)$, canonical
coordinate and momentum,  by
\begin{align}
  q=&\frac{1}{\sqrt{2}}\frac{\mu+\mu^*}{\sqrt{1+|\mu|^2}}~, \\
  p=&\frac{-i}{\sqrt{2}}\frac{\left(\mu-\mu^*\right)}{\sqrt{1+|\mu|^2}}~. 
\end{align}
Applying this transformation, the resulting classical Hamiltonian is
\begin{equation}\label{Hqp}
  H_{cl}(q,p)=\frac{1-\xi}{2}\left(p^2+q^2\right)+\frac{\alpha}{4}\left(p^2+q^2\right)^2+\xi q^2\left(p^2+q^2-2\right) + \xi~.
\end{equation}

The resulting energy functional should  provide the functional form of the separatrices that mark the critical ESQPT energies in  Fig.~\ref{ced_figure}, one in the $\alpha = 0$ case (upper panel) and two for $\alpha < 0$  (lower panel). The separatrix starting at the critical value of the control parameter $\xi_c=0.2$ has been already characterized for $\alpha=0$. We present results for both ESQPTs with a special focus on the anharmonicity-related critical line; exploring whether  this line has the same nature and implications than the other and to what extent
it has an impact in the system structure. The changes in the system dynamics  associated with the introduction of the anharmonicity in the LMG model are explored in a accompanying publication \cite{Gamito2022}.

Assuming $\xi\in[0,1] $ and $\alpha<0$, the critical or stationary points of the Hamiltonian, where the first
derivatives of  Eq.~\eqref{Hqp} are zero, are
\begin{equation}
  \begin{array}{cc}
    \pdv{H_{cl}(q,p)}{q} & = 0 \\
    \pdv{H_{cl}(q,p)}{p} & = 0
  \end{array}
  \longrightarrow
  \left\{
    \begin{array}{ll}
     &  \left(q_0^2=0, p_0^2=0\right) \\
     &  \left(q_1^2=\frac{5\xi-1}{4\xi+\alpha}, p_1^2=0\right) \\
      &  \left(q_2^2=\frac{\xi-1-2\alpha}{2\xi}, p_2^2=\frac{1+2\alpha+3\xi}{2\xi}\right) \\
     &  \left(q_3^2=0, p_3^2=\frac{\xi-1}{\alpha}\right) ~~.
    \end{array}
  \right. 
\end{equation}
According to the classical limit, the origin is a stationary point
and it corresponds with the system ground state in the control parameter
range $\xi \in [0, \xi_c]$ where the critical point is $\xi_c = 0.2$. For values of the control
parameter $\xi \in (\xi_c, 1]$, the minimum energy is attained for $(q_1,p_1)$ coordinate and momentum values. In this way, the system ground state energy as a function of the control parameters can be expressed as follows
  \begin{equation}\label{Egs2nd}
  E_{gs}(\xi,\alpha) = \left\{
    \begin{array}{ll}
      \xi &; \xi\leq 0.2 \\
      \frac{-1 + \xi\left(10+4\alpha-9\xi\right)}{4\left(\alpha+4\xi\right)} &; \xi > 0.2
    \end{array}
  \right.~~.
\end{equation}
From this equation is clear that the Hamiltonian Eq.~\eqref{eq:LMG2}
in the mean field limit has a critical point at $\xi_c = 0.2$, where
the second derivative of the ground state energy with respect to the
control parameter $\xi$ is discontinuous. This is in good agreement
with the results obtained in \cite{Cejnar2007} for a Hamiltonian
including one- and two-body operators. Hence, the crossing of this
critical point is marked by a second order ground state QPT. The
energy gap vanishes and the ground state becomes parity
degenerated. In addition to this, and in a similar way to other
systems with a transition between $u(n)$ and $so(n+1)$ dynamical
symmetries \cite{PBernal2008}, an ESQPT appears at an energy
$H_{cl}(q_0,p_0)=\xi$. The critical excitation energy of the ESQPT
defines the separatrix with a high density of states shown in the
upper and lower panels of Fig.~\ref{ced_figure}. An analytical
expression for this separatrix can be computed as follows
\begin{equation}\label{f1}
  f_1(\xi,\alpha) = H_{cl}(q_0,p_0) - H_{cl}(q_1,p_1) = \frac{ (1-5\xi)^2 }{4(4\xi+\alpha)}~,
\end{equation}
\noindent for $\xi \in (\xi_c, 1]$.

For negative $\alpha$ values there can be a second separatrix in this system, as shown in the lower panel of Fig.~\ref{ced_figure}.  When the condition $\alpha<(\xi-1)/2$ is satisfied, the second ESQPT
appears at a critical energy $H_{cl}(q_2,p_2)=1+\alpha$. The excitation energy for this new ESQPT marks the second separatrix in the lower panel of Fig.~\ref{ced_figure} and can be computed as
\begin{equation}
  f_2(\xi,\alpha) = 
  \left\{
    \begin{matrix}
      H_{cl}(q_2, p_2) - H_{cl}(q_0,p_0) & = & 1+\alpha-\xi & \xi\leq \xi_{c} \\
      H_{cl}(q_2, p_2) - H_{cl}(q_1,p_1) & = & \frac{(1+2\alpha+3\xi)^2}{4(\alpha+4\xi)} & \xi >\xi_{c}
    \end{matrix}~.
  \right.\label{f2}
\end{equation}
As previously mentioned, a similar ESQPT in the broken symmetry phase of the 2D limit of the vibron model was studied in Ref.~\cite{PBernal2010} and it has recently been shown to be present in the symmetric phase of the model too \cite{KRivera2021}.

In the symmetric phase, $\xi\in[0,\xi_c]$, pair of eigenstates with
different parities are degenerate for energies above the
$f_2(\xi,\alpha)$ separatrix. In the broken-symmetry phase,
$\xi\in(\xi_c,1]$, both separatrices can coexist, and the pairs of
eigenvalues with even and odd parities are degenerate below and above
both of them and non-degenerate in between the two separatrices. This
is clearly evinced in the lower panel of Fig.~\ref{ced_figure}. The
critical point $(q_3,p_3)$ marks the largest energy of the system ,
$H_{cl}(q_3,p_3)=\frac{-1+(2+4\alpha-\xi)\xi}{4\alpha}$, if the
solution $(q_2,p_2)$ exists. In fact, taking this into account and
equating $H_{cl}(q_2,p_2) = H_{cl}(q_3,p_3)$, the threshold value of
$\alpha$ in order that the second ESQPT exists can be obtained as
$\alpha_{th}(\xi) = (\xi -1)/2$. From this equation it is clear that
the second ESQPT is not only present in the broken-symmetry phase, but
also in the symmetric phase. This was recently discussed for a
different model, the 2DVM \cite{KRivera2021} and it is a clear example
of an ESQPT without an associated QPT, something that usually happens
in systems with more than one control parameter for some trajectories
in the parameter space \cite{Relano2016, Stransky2021,gatoAngel}. In
this case, the situation is somewhat different and the second ESQPT
can be associated to changes in the boundaries in the finite Hilbert
space of the system \cite{BMagnani2014, Macek2019, Cejnar2021}.

\begin{figure}
    \centering
    \subfigure[]{
    \includegraphics[width=1.0\textwidth]{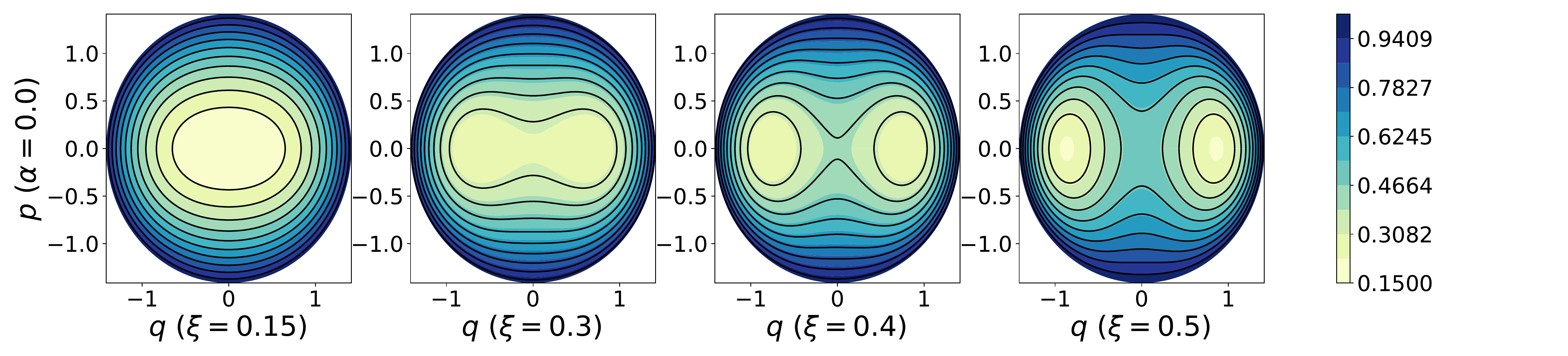}
    }
    
    \subfigure[]{
    \includegraphics[width=1.0\textwidth]{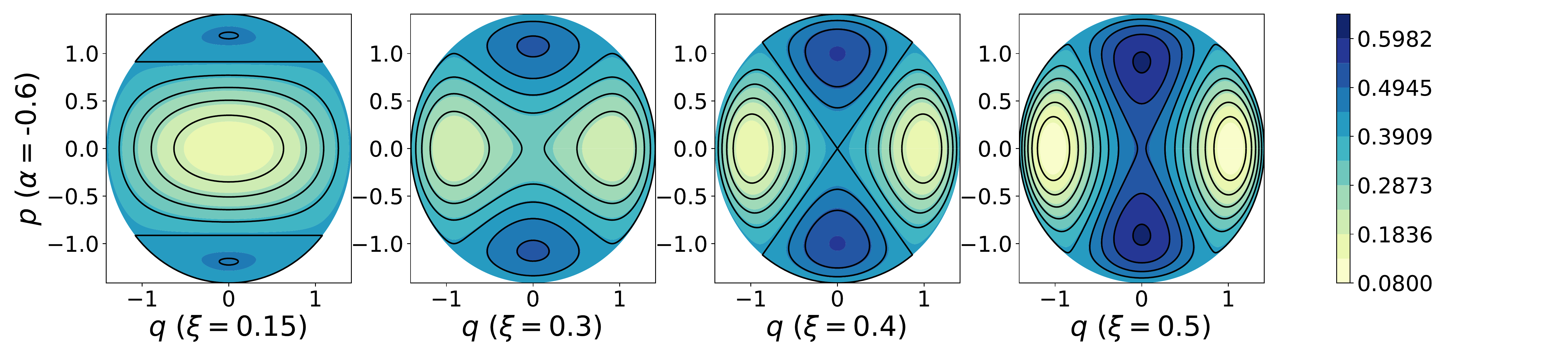}
    }
    \caption{Contour plots of the classical Hamiltonian \eqref{Hqp} for systems with $\alpha=0.0$ (upper row) and $-0.6$ (lower row), and different values of the control parameter, from left to right $\xi=0.15$, $0.3$, $0.4$ and $0.5$. Each row shares the scale (right color axes).
    \label{fig:contourH}}
\end{figure}

In Fig.~\ref{fig:contourH} we show the contour plot of the classical Hamiltonian \eqref{Hqp} for two different values of the anharmonic interaction, one without the $f_2$ (Eq. (\ref{f2}) ESQPT, $\alpha=0.0$ (upper row), and another with a value of the control parameter smaller than the threshold, $\alpha=-0.6$ (lower row). In both cases different values of the control parameter $\xi$ have been studied, from left to right $\xi=0.15$, $0.3$, $0.4$ and $0.5$. The cases with $\alpha=0.0$ present one minimum for $\xi<0.2$ and two minima in the broken-symmetry phase (upper row of Fig~\ref{fig:contourH}). In the four examples the maximum values of the energy correspond to $q_2^2+p_2^2=2$. When the control parameter $\alpha$ takes a value smaller than the threshold (lower row of Fig.~\ref{fig:contourH}), there appear two maxima for $q_3=0$ and $p_3\neq 0$, which correspond to the new system limit. In these cases, systems present a new ESQPT in due to the system limit $q_2^2+p_2^2=2$.

\begin{figure}
\begin{centering}
  \includegraphics[scale=0.5]{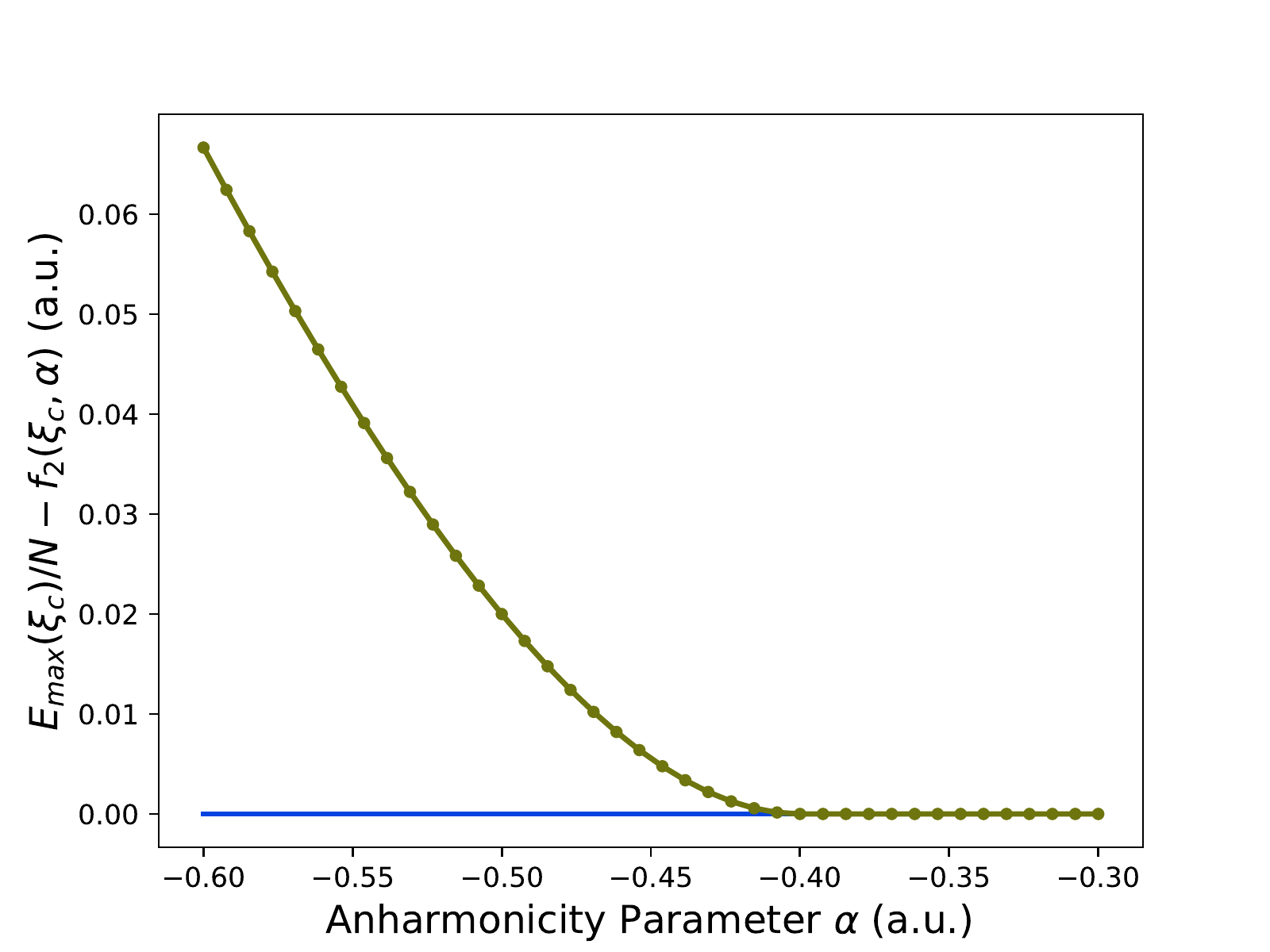}
  \par\end{centering}
\caption{(Color online) The full green olive line is the difference between the maximum normalized excitation energy of the aLMG (see text) and the separatrix Eq.~\eqref{f2} for \(\xi = \xi_c = 0.2\). The blue horizontal line marks the zero value.}
\label{alpha_thresh_fig}
\end{figure}

In order to check the threshold value $\alpha_{th}(\xi)$, we have computed the maximum excitation
energy for different $\alpha$ values and system sizes and we have fit the maximum normalized excitation energy to a functional form $E_{max} = E_{cl} + c N^{-b}$ for a constant $\xi = \xi_c$. The
resulting $E_{cl}$ values for $\xi_c$ and various $\alpha$ values are
a good estimate of the maximum excitation energy of the system in the
classical limit. The difference between $E_{cl}$ and the
separatrix $f_2(\xi_c,\alpha)$ is depicted in Fig.~\ref{alpha_thresh_fig} where it is
clear how the threshold value of $\alpha$ exists and is equal to the value $\alpha_{th}(\xi_c) =-0.4$.

As can be clearly seen in Fig.~\ref{ced_figure}, the existence of an ESQPT can be traced back to abrupt changes in the system available phase space which produces a maximum in the local density of states at the separatrices, for the critical energy of the transition \cite{PFernandez2009, BMagnani2014,Stransky2014,Cejnar2021}. From the classical Hamiltonian Eq.~\eqref{Hqp}, we can evaluate the semiclassical approximation to the quantum density of states as a function of the system's energy \cite{Gutzwiller2013}. In the particular case of the LMG and aLMG models 
\begin{equation}
  \nu(E) = \frac{1}{2\pi}\int\int dq dp ~\delta(E-H_{cl}(q, p))= \frac{1}{2\pi}\int\int d\phi d j_z ~\delta(E-H_{cl}(\phi, j_z))~,
  \label{doseq}
\end{equation}
\noindent where instead of the phase space generalized coordinate and associated momentum we used a different pair of canonical variables, $\phi\in [0, 2\pi)$ and $j_z\in [-j, j]$ that facilitate the derivation of an analytical formula for the density of states of the aLMG model. Using the properties of the Dirac delta function (see App.~\ref{appa}), one can obtain the analytical formula for the density of states that follows
  \begin{equation}
    \nu(\varepsilon) = \frac{1}{4\pi}\bigintss_0^{2\pi}{\frac{d\phi}{\left|\sqrt{\left(\frac{1-\xi+\alpha}{2}\right)^2-(\alpha + 4\xi\cos^2\phi)\left(\frac{2+2\xi+\alpha}{4}-\xi\cos^2\phi - \varepsilon\right)}\right|} }~,\label{analdos}
  \end{equation}
  \noindent where $\varepsilon$ is the scaled energy value $\varepsilon = E/N$.
  
\begin{figure}
\begin{centering}
  \includegraphics[scale=0.4]{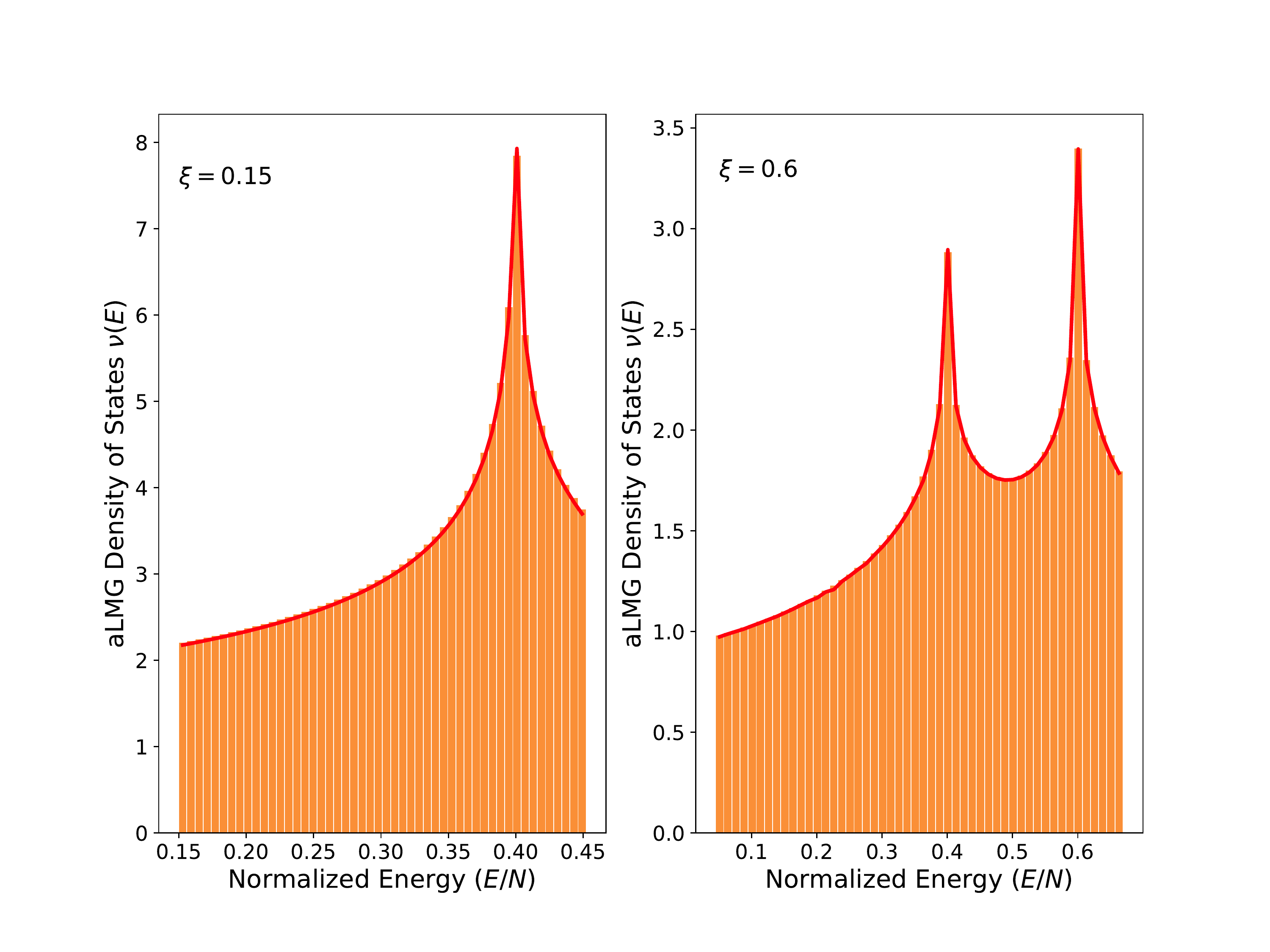}
  \par\end{centering}
\caption{(Color online) Density of states $\nu(E)$ as a function of the normalized energy $E/N$ for an aLMG Hamiltonian  $\alpha = -0.6$ and $\xi = 0.15, 0.6$. The full bright red line is the $\nu(E)$ density of states computed with Eq.~\eqref{doseq}. We have also included the density of states computed for the normalized eigenvalues of an aLMG Hamiltonian Eq.~\eqref{eq:LMG2} with $N = 2000$ (bright orange bars).}
\label{alpha_dos}
\end{figure}

In Fig.~\ref{alpha_dos} we plot the density of states $\nu(E)$ calculated with Eq.~\eqref{analdos} (red line) versus the normalized energy for $\alpha = -0.6$ (the same $\alpha$ value used in the correlation energy diagrams of Fig.~\ref{ced_figure}) and two $\xi$ values: $0.15$ (symmetric phase, left panel) and $0.6$ (broken-symmetry phase, right panel). For the sake of comparison, we also include the density of states computed from the eigenvalues of an aLMG  Hamiltonian \eqref{eq:LMG2} for the same control parameters and a system's size $N = 2000$. It can be clearly appreciated  how the density of states in the left panel has a maximum at the $f_2(\xi = 0.15, \alpha = -0.6)+E_{gs}(\xi = 0.15, \alpha = -0.6)=0.4$  critical energy  in the symmetric phase  and in the right panel at $f_1(\xi = 0.6, \alpha = -0.6)+E_{gs}(\xi = 0.6, \alpha = -0.6)=0.4$ (leftmost maximum) and  $f_2(\xi = 0.6, \alpha = -0.6)+E_{gs}(\xi = 0.6, \alpha = -0.6)=0.6$ (rightmost maximum) critical energies in the broken symmetry phase. The agreement between the results for Eq.~\eqref{analdos} and the density of states computed from the system energies is excellent and it is worth to notice how in all cases the peaks will tranform into logarithmic divergences for an infinite system size, as expected \cite{Stransky2016}. This has deep implications in the system dynamics, as it is  evinced in an associated work \cite{Gamito2022}. 

In the next section we characterize the two ESQPTs and its associated separatrices using different quantities.

\section{Results}
\label{secIII}

In this section we characterize the two ESQPTs that arise in the aLMG using
four different quantities: the gap between adjacent positive parity energy levels, the
expectation value of the operator $\hat n = S+\spin_{z}$, the participation
ratio, and the quantum fidelity susceptibility.

\bigskip

\paragraph{Energy gap} The occurrence of an ESQPT is most often marked by a
discontinuity in the density of states or one of its derivatives. In
cases such as the LMG model \eqref{eq:LMG1}, for a given value of the control
parameter $\xi > \xi_c = 0.2$, the difference in energy between adjacent eigenstates is
minimum at or close to the critical energy of the ESQPT, due to the high density
of states at this excitation energy (see Fig.~\ref{ced_figure}). Results are more complex once we introduce the anharmonic correction in the LMG model Hamiltonian Eq.~\eqref{eq:LMG2}.  

\begin{figure}
\begin{centering}
\includegraphics[scale=0.5]{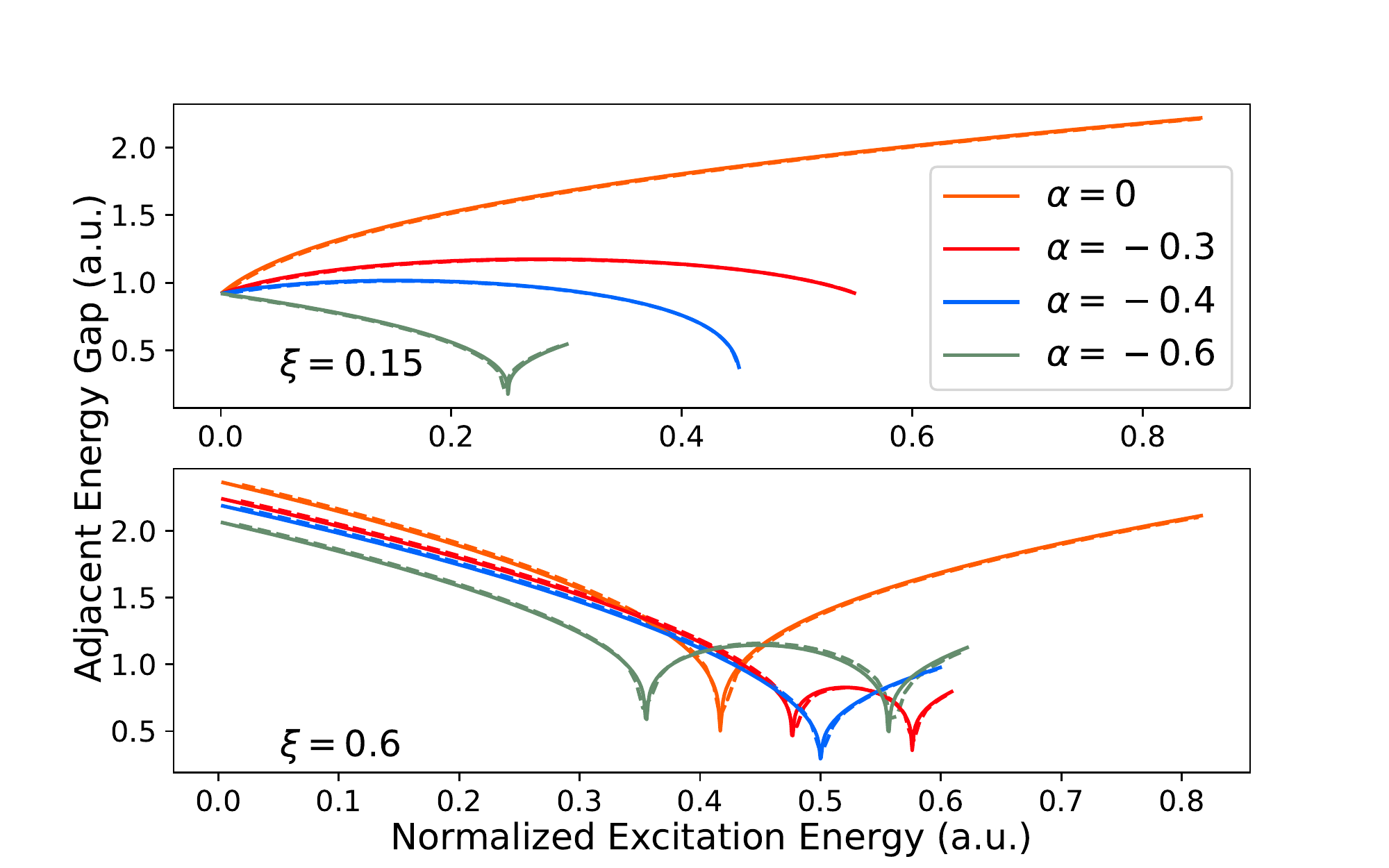}
\par\end{centering}
\caption{(Color online) Gap between adjacent energy levels in the aLMG model as a function of the normalized excitation energy for $\xi = 0.15$ (upper panel) and $0.6$ (lower panel) and different values of the anharmonicity parameter $\alpha$ (see legend) and system sizes $N = 120$ (dashed lines) and $1200$ (full lines).}
\label{neg_fig}
\end{figure}

In Fig.~\ref{neg_fig} we plot the difference between adjacent positive parity energy levels as a function of the normalized excitation energy for system size values $N = 2S = 120$ (dashed lines) and $1200$ (full lines), for control parameter $\xi$ equals to $0.15$ (upper panel) and $0.6$ (lower panel), and $\alpha = 0, -0.3, -0.4, -0.6$. 

The results in the upper panel, where the system is in the symmetric phase ($\xi = 0.15 < \xi_c$),  can be explained considering the second separatrix Eq.~\eqref{f2}. The $f_2(\xi,\alpha)$ line  marks the critical energy value for the ESQPT associated with the anharmonic term in $\xi > \xi_c$ and, as explained above, there is a threshold $\alpha$ and the effects of the anharmonic term in this phase can be noticed only when the $\alpha$ parameter is beyond this threshold value (see green line in upper panel Fig.~\ref{neg_fig}). The crossing of the critical energy is marked by an abrupt minimum in the energy gap, as expected. When this study is extended to the broken-symmetry phase (lower panel in  Fig.~\ref{neg_fig}) there are two minima, one for each separatrix, for negative $\alpha$ values except for $\alpha = -0.4$ as this corresponds to the point associated with the crossing of $f_1(\xi,\alpha)$ and $f_2(\xi,\alpha)$ in the lower panel of Fig.~\ref{ced_figure}. As expected, the ESQPT precursor is better defined the larger the system size, though for $N= 2s = 120$ the results already clearly identify the ESQPT. The aLMG energy gap results are in good agreement with the 2DVM results obtained in the symmetric \cite{KRivera2021} and broken-symmetry \cite{PBernal2010} cases.

\bigskip

\begin{figure}
\begin{centering}
\includegraphics[scale=0.5]{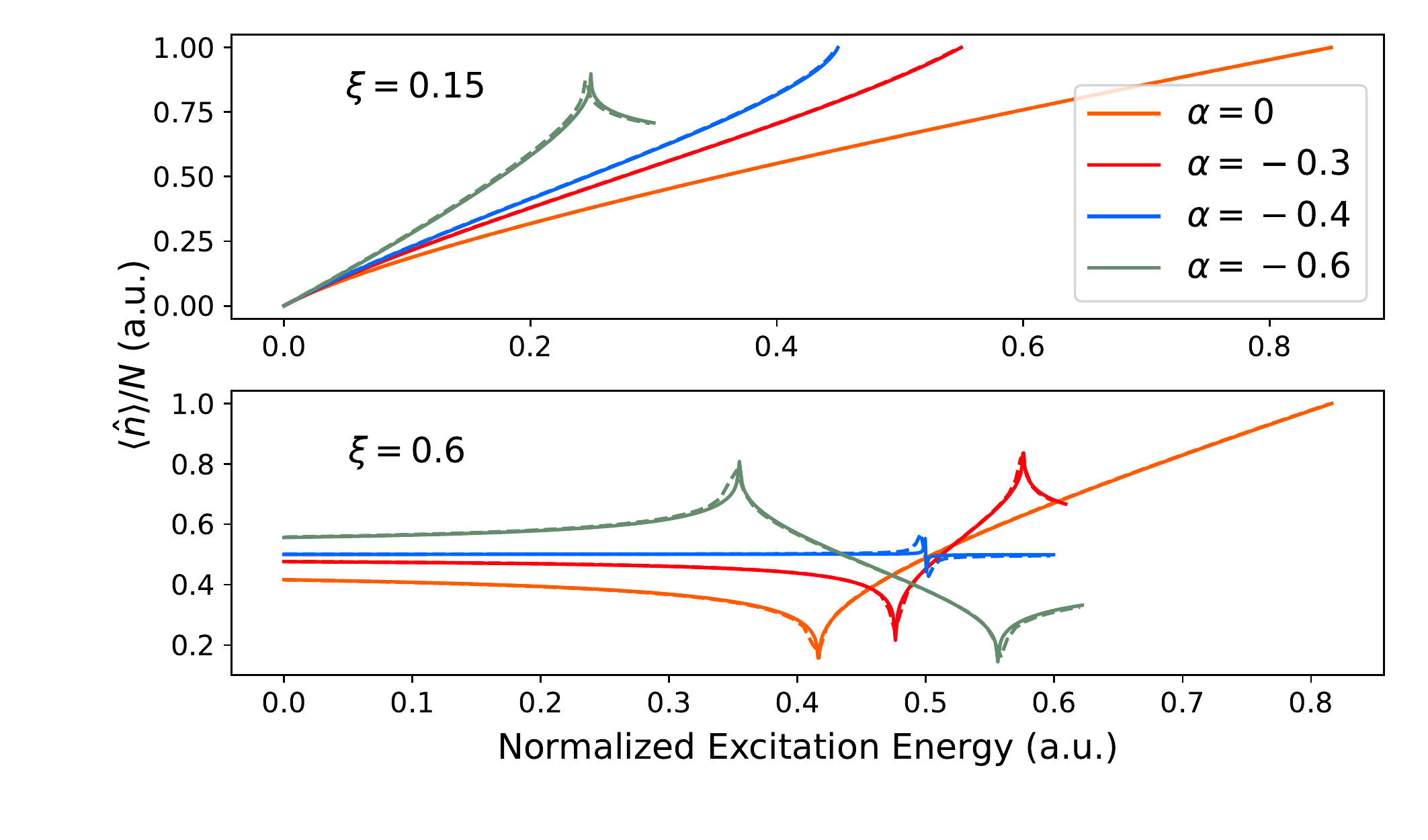}
\par\end{centering}
\caption{Expectation value of the number operator $\hat n$ in the system eigenstates for $\xi = 0.15$ (upper panel) and $0.6$ (lower panel), different values of the anharmonicity parameter $\alpha$ (see legend), and system sizes $N = 120$ (dashed lines) and $1200$ (full lines).}
\label{expn_fig}
\end{figure}

\paragraph{Number operator}
The $S+\spin_{z}$ term in Eqs.~\eqref{eq:LMG1} and \eqref{eq:LMG2} is associated with the $u(1)$ Casimir operator and it can be denoted as the system number operator, $\hat n$. This Casimir operator provides a suitable realization of an order parameter for the ground state QPT in the aLMG case \cite{Botet1983, Dusuel2004, Dusuel2005,Ribeiro2008,Santos2016}. The expectation value of this operator in the system eigenstates is a second ESQPT proxy in Hamiltonian  Eq.~\eqref{eq:LMG2}. We depict in Fig.~\ref{expn_fig} the expectation value of  $\hat n$ as a function of the normalized system excitation energy for  $\xi = 0.15$ (upper panel) and $0.6$ (lower panel), and different values of the anharmonicity parameter $\alpha$ (see legend) for system sizes $N = 120$ (dashed lines) and $1200$ (full lines). In all cases, the expectation value of $\hat n$ is a well defined minimum (maximum) for the eigenstates closer to the $f_1(\xi,\alpha)$ ($f_2(\xi,\alpha)$) separatrix. This fact will be better understood once we introduce the inverse participation ratio to characterize both ESQPTs.

It is worth to emphasize that in the  symmetric case there is no minima, the $f_1(\xi,\alpha)$ line is only defined for $\xi > \xi_c$. This is clearly shown in the upper panel of  Fig.~\ref{expn_fig}, where $\xi = 0.15<\xi_c$. And, as in the energy gap case, this quantity has a peak only for values beyond the $\alpha$ threshold value that corresponds to the second separatrix,  $f_2(\xi,\alpha)$.

\bigskip

\begin{figure}
\begin{centering}
\includegraphics[scale=0.65]{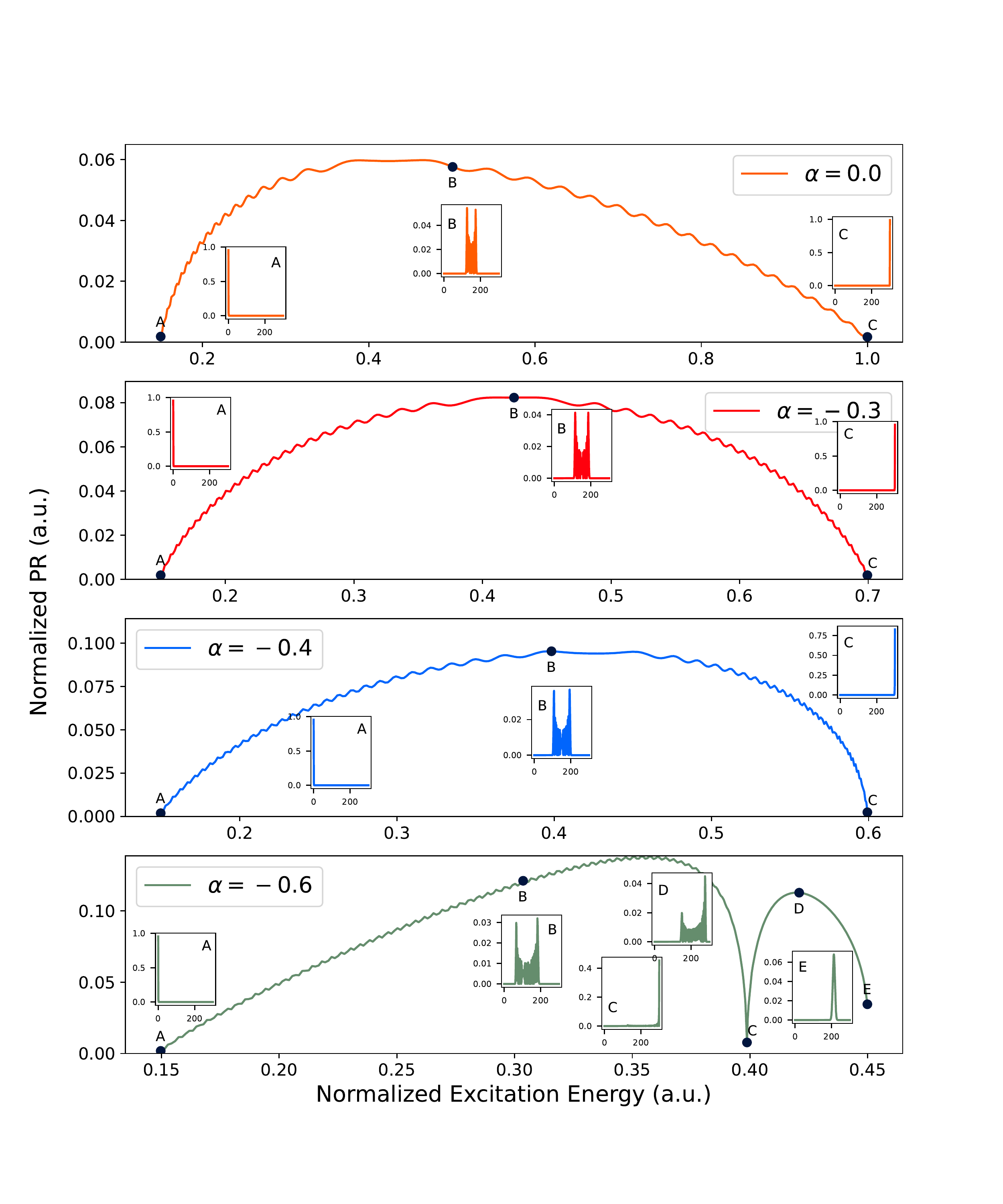}
\par\end{centering}
\caption{Normalized participation ratio as a function of the normalized excitation energy for an aLMG model with $\xi = 0.15$ and system size $N = 600$. The four panels correspond to different anharmonicity parameter values: $\alpha = 0, -0.3, -0.4, -0.6$. In each case, the PR values for eigenstates of interest have been marked with black dots and the squared components of this eigenstates as a function of a corresponding $u(1)$ basis state index is shown in the insets (see text). }
\label{ipr_fig_015}
\end{figure}

\begin{figure}
\begin{centering}
\includegraphics[scale=0.65]{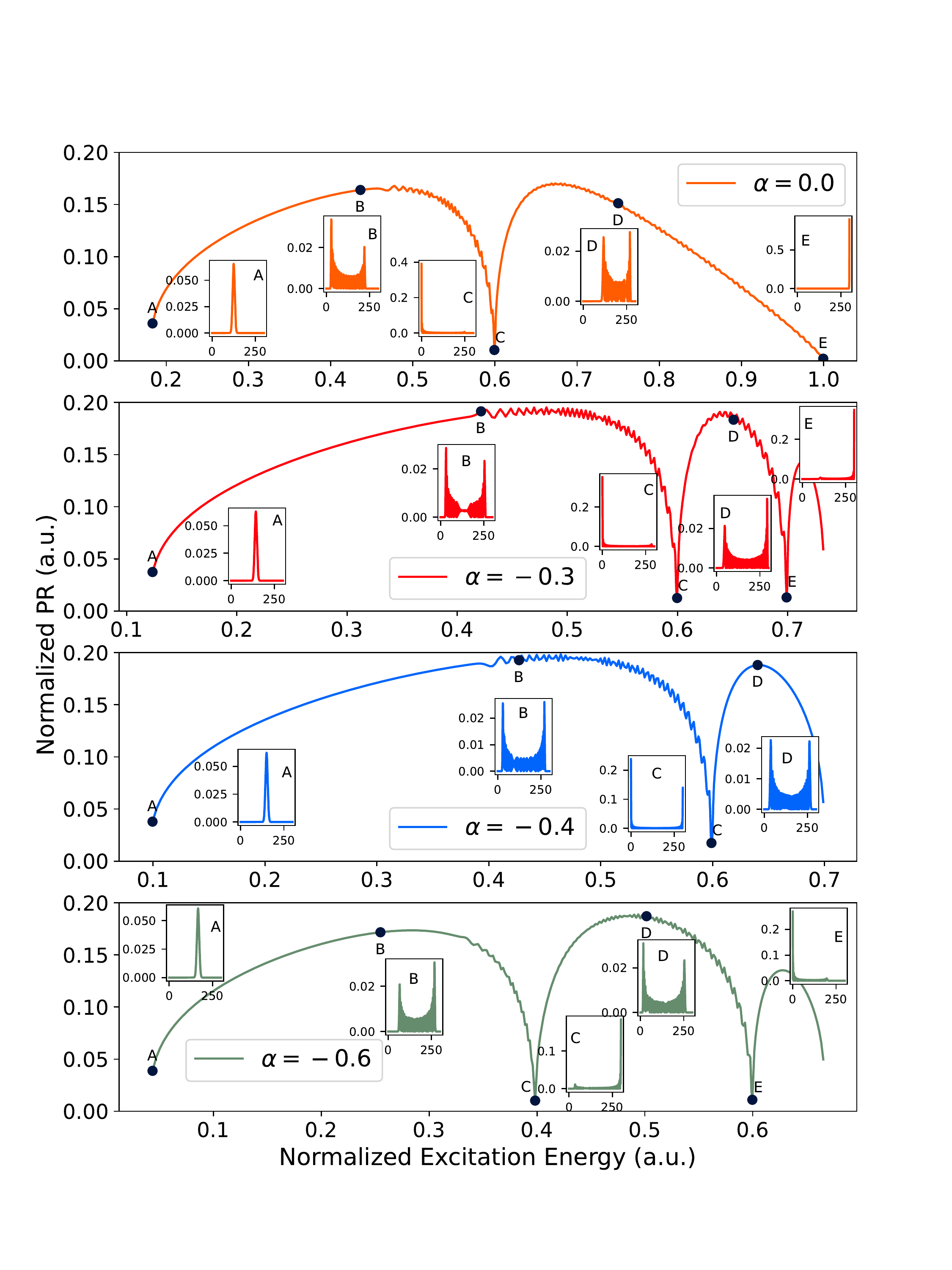}
\par\end{centering}
\caption{Normalized participation ratio as a function of the normalized excitation energy for an aLMG model with $\xi = 0.6$ and system size $N = 600$. The four panels correspond to different anharmonicity parameter values: $\alpha = 0, -0.3, -0.4, -0.6$. In each case, the PR values for eigenstates of interest have been marked with black dots and the squared components of this eigenstates as a function of a corresponding $u(1)$ basis state index is shown in the insets (see text). }
\label{ipr_fig_06}
\end{figure}

\paragraph{Participation ratio}
The participation ratio (PR) provides the localization of a quantum state in a given basis \cite{Izrailev1990}. For a state \(\ket{\psi}\), expressed in a basis \(\{\ket{n}\}_{n = 0}^{D-1}\) as \(\ket{\psi}= \sum\limits_{n=0}^{D-1} C_{n} \ket{n} \), the PR is defined as 
\begin{equation}
  P\left(\psi\right)=\frac{1}{\sum\limits_{n=0}^{D-1} \left|C_{n}\right|^4}~.
\end{equation}
\noindent Where $D$ stands for the Hilbert space's dimension, that in this case is $D=N+1$. This quantity that is also often denoted as inverse participation ratio \cite{Evers2008} or number of principal components \cite{Zelevinsky1996}. In case that the state is maximally delocalized, $C_n = 1/\sqrt{D}$ for all $n$ and $P\left(\psi\right) = D$; if the state is equal to one basis element (all $C_n = 0$ but one that is equal to unity) the localization is maximal in the basis and $P\left(\psi\right) = 1$. A series of works in algebraic models for systems in one-, two-, and three-dimensions having a $u(n+1)$ dynamical algebra has shown that the eigenstates closer to the critical energy in a $u(n)-so(n+1)$ second order ground state quantum phase transition are strongly localized in the $u(n)$ basis~\cite{Santos2015, Santos2016, PBernal2017}. In the aLMG case there is a strong localization in the $u(1)$ basis of the states close to either one of the two separatrices. In particular, the eigenstates with energies close to the critical energy of the $f_1(\xi,\alpha)$ separatrix are highly localized in the $n=0$ component ($\ket{S,M_z=-S}$) of the basis, something that was already checked in \cite{Santos2016} for the LMG model. However, the localization of the eigenstates with energies close of the second separatrix, $f_2(\xi,\alpha)$, critical energy is due to a high component in the  $n=N$ basis state ($\ket{S,M_z=S}$). This is evinced in Figs.~\ref{ipr_fig_015} and \ref{ipr_fig_06} for systems  in the symmetric phase ($\xi = 0.15$) and in the broken-symmetry phase, respectively. In both cases, the system size is $N= 600$ and the four vertical panels depict results for  $\alpha = 0, -0.3, -0.4$, and $-0.6$. In each panel some representative eigenstates have been chosen and the value of their squared components in the $u(1)$ basis are shown in the corresponding inset.

In Fig.~\ref{ipr_fig_015} the only $\alpha$ value beyond the theshold value is $\alpha = -0.6$, shown in the lower panel. In the three upper panels the PR is minimum at the spectrum edges, where eigenstates are maximally localized in the $\ket{S,M_z=-S}$ and $\ket{S,M_z=S}$ states, as expected in the symmetric phase where the $u(1)$ dynamical symmetry provides a convenient approximation. However, in the lower panel (olive color line), a well-defined PR minimum appears associated to the  $f_2(\xi,\alpha)$ separatrix and the ESQPT induced by the anharmonic term in Eq.~\eqref{eq:LMG2}. The eigenstate labeled as C in this panel -- the eigenstate with a minimum PR value which is not in the spectrum edges -- is highly localized (low PR) and its squared components in the inset reveal that the localization takes place for the last basis state, $\ket{S,M_z=S}$.

The PR values in the broken-symmetry phase are shown in Fig.~\ref{ipr_fig_06}. In the $\alpha = 0$ case (upper panel, orange curve), the states at the edge of the spectrum have low values of the PR, as expected, and the ground state is not well localized in the $n=0$ basis state, as we have straddled the critical ground state QPT  and the $u(1)$ dynamical symmetry is not anymore  a convenient approximation for the system eigenstates. There is a local PR minimum, labeled as C, associated with the $f_1(\xi,\alpha)$ separatrix and the ESQPT that stems from the appearance of a maximum in the origin. When $\alpha = -0.3$ (second panel, red curve), precursors from both ESQPTs can be observed. In this case the $f_1(\xi,\alpha)$ separatrix is found  at lower energies than the $f_2(\xi,\alpha)$ separatrix and the states close to the separatrices and that constitute local minima for the PR are labeled as C and E, respectively. Both states are strongly localized and, from the information in the insets, the first one has a dominant component in  the $\ket{S,M_z=-S}$ basis state, and the second case in the $\ket{S,M_z=S}$ basis state. The situation is reversed in the  $\alpha = -0.6$ case (lower panel, olive curve) as the anharmonicity induced ESQPT (C state) lies now at lower energies than the original ESQPT (E state). The third panel (blue curve) is a special case, where only one PR local minimum appears, besides the spectrum edges,  as for this particular $\xi$ and $\alpha$ values there is a crossing of the two separatrices. The eigenstate having a local minimum value of the PR, labeled as C, is well localized with high components in both $\ket{S,M_z=-S}$ and $\ket{S,M_z=S}$ basis states. This is in good accordance with results obtained for an anharmonic  2DVM  Hamiltonian \cite{KRivera2019, KRivera2021}.

\bigskip 

\paragraph{Quantum fidelity susceptibility}
Quantum fidelity for a system with a single control parameter, $\lambda$, and a Hamiltonian \(\hat H(\lambda) = \hat H^0 + \lambda \hat H^I\), is defined as the modulus of the overlap between the ground state of the system for  $\lambda$ and $\lambda + \delta\lambda$
\begin{equation}
  F(\lambda,\delta\lambda) =
  \left|\bra{\psi_0(\lambda)}\ket{\psi_0(\lambda + \delta\lambda)}\right|~.
\end{equation}
This quantity, originally introduced in the realm of quantum information \cite{Nielsen2000}, was later used for the characterization of QPTs as it efficiently tracks the sudden change in the wave function of the system ground state as it straddles the critical value of the control parameter \cite{Zanardi2006, You2007, Gu2010}. The quantum fidelity susceptibility (QFS) is defined as the second order -and leading- term
in the series expansion of  $F(\lambda,\delta\lambda)$ as a function of
\(\delta\lambda\) \cite{You2007,Gu2010}. 
\begin{equation}
  \chi_{F}(\lambda)= -\frac{\partial^2F(\lambda,\delta\lambda)}{\partial(\delta\lambda)^2} = \lim_{\delta\lambda\to 0}\frac{-2\ln{F(\lambda,\delta\lambda)}}{(\delta\lambda)^2}~.\label{chi_F}
\end{equation}
The QFS reaches a maximum at the critical value of the \(\lambda\) control parameter and it is
independent of the \(\delta\lambda\) value. Using first-order perturbation theory, the QFS can be expressed as \cite{Gu2010}
\begin{equation}
\chi_{F}(\lambda)= \sum^{D-1}_{i\ne0}\frac{\left|\bra{\psi_i(\lambda)} \hat H^I\ket{\psi_0(\lambda)}\right|^2}{\left[E_i(\lambda)-E_0(\lambda)\right]^2}~~,\label{fidsuscep}
\end{equation}
\noindent where \(\ket{\psi_i(\lambda)}\) is the \(i\)-th eigenvector of the Hamiltonian \(\hat
H(\lambda)\) and \(E_i(\lambda)\) is its associated eigenvalue.

Recently, some of the authors (JKR and FPB) have suggested to extend the use of the QFS to the study of ESQPTs using as an example the 2DVM and its application to the bending of nonrigid molecules \cite{KRivera2022}. Following the proceduere detailed in this reference, we define a Hamiltonian with a new control parameter, $\lambda\in[-1,1]$, grouping the interactions in Eq.~\eqref{eq:LMG2} considering the dynamical symmetry they belong to

\begin{eqnarray}
\hat H(\lambda)&=&(1-\lambda)\left\{(1-\xi)\left(S+\spin_{z}\right)+\frac{\alpha}{2S}\left(S+\spin_z\right)\left(S+\spin_z+1\right)\right\} + (1+\lambda)\left\{\frac{2\xi}{S}\left(S^2-\spin_{x}^2\right)\right\}~,\\
\label{eq:LMG_Lambda}
\hat H(\lambda)&=& \hat H_0 + \lambda \hat H^I~,~ \hat H^I= -(1-\xi)\left(S+\spin_{z}\right) - \frac{\alpha}{2S}\left(S+\spin_z\right)\left(S+\spin_z+1\right) + \frac{2\xi}{S}\left(S^2-\spin_{x}^2\right)~.
\end{eqnarray}
In this way, if our starting point is a Hamiltonian  Eq.~\eqref{eq:LMG2} with given values of $\xi$ and $\alpha$, the Hamiltonian Eq.~\eqref{eq:LMG_Lambda} for $\lambda = -1$($\lambda = 1$) is diagonal in the $u(1)$($so(2)$) basis and we recover the original Hamiltonian for $\lambda = 0$.

The QFS definition in Eq.~\eqref{fidsuscep} can be extended to encompass excited states of Hamiltonian  \eqref{eq:LMG_Lambda} \cite{KRivera2022}
\begin{equation}
\chi^{(j)}_{F}(\lambda)= \sum^{D-1}_{i\ne j}\frac{\left|\bra{\psi_i(\lambda)} \hat H^I\ket{\psi_j(\lambda)}\right|^2}{\left[E_i(\lambda)-E_j(\lambda)\right]^2}~,\label{fidsuscepII}
\end{equation}
\noindent where $\ket{\psi_j(\lambda)}$ is the $j$-th eigenstate of the $ \hat H(\lambda)$ Hamiltonian with $j = 0, \ldots, D-1$.

\begin{figure}
    \centering
    \includegraphics[width=1.0\textwidth]{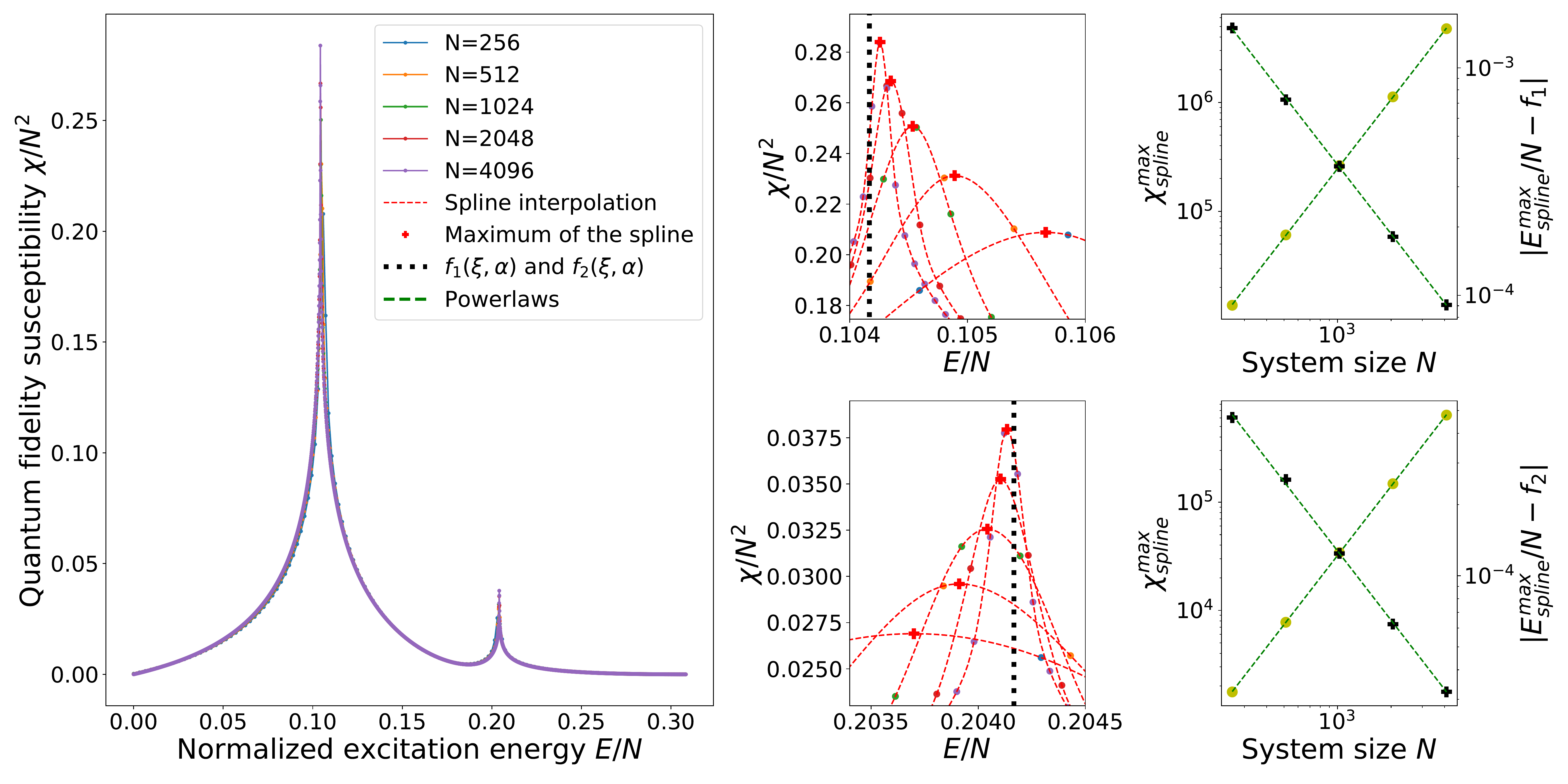}
    \caption{Normalized quantum fidelity susceptibility $\chi^{(j)}_{F}(\lambda =0)/N^2$ as a function of the normalized excitation energy for the aLMG with $\xi=0.3$ and $\alpha=-0.6$ (main panel) for different system sizes $N=256$ (blue), $512$ (orange), $1024$ (green), $2048$ (red), and $4096$ (purple). On the right of the figure there are four additional panels. The first column corresponds to the zoom of each extreme of the QFS. The mean field values of the critical energies -eqs. \eqref{f1} and \eqref{f2}- have been plotted with a black dotted line. The red dashed lines correspond to the interpolation with splines used to locate the positions of maxima, which have been highlighted with red crosses. In the second column we show the maximum values obtained with the interpolation of the QFS $\chi^{\text{max}}_{\text{spline}}$ (gold points and left axes) and the distances of the energies where these maxima occur to the mean field critical energies $\left|E^{\text{max}}_{\text{spline}}/N-f_{\text{mf}}\right|$ (black crosses and right axes) versus the system size $N$ using log scales. The power interpolation have been added (green dashed lines) in all cases. See text for details.}
    \label{fig:QFS}
\end{figure}

In Fig.~\ref{fig:QFS} we report the normalized QFS $\chi/N^2$ as a function of the normalized excitation energy $E/N$ for the aLMG model with $\xi=0.3$ and $\alpha=-0.6$, system which is located between the ground-state QPT and the separatrices cross, and different system sizes $N=256$ (blue), $512$ (orange), $1024$ (green), $2048$ (red), and $4096$ (purple). In both transitions the QFS is maximum near the critical energy, being the maximum due to $f_1$ (lower energy) higher than the $f_2$ one (higher energy). In the first column of the four additional panels, we show a detailed zoom of peaks, lower energy transition (upper panel) and higher energy transition (lower panel). We have added an interpolation with splines for each system (red dashed lines) and highlighted its maximum value (red crosses). Therefore, the mean field values of the critical energies have been plotted with black dotted lines. In the second column we plot the maximum value of QFS $\chi^{\text{max}}_{\text{spline}}$ (gold points and left axes) as well as its position respect to the mean field value $\left|E^{\text{max}}_{\text{spline}}/N-f_{\text{mf}}\right|$ (black crosses and right axes) using log scales. As expected, in both transitions the critical energy tends to the mean field value  and the QFS diverges according to power laws,
\begin{align}
    \chi^{\text{max}}_{\text{spline}} & \propto N^{a} \\
    \left|E^{\text{max}}_{\text{spline}}/N-f_{\text{mf}}\right| &\propto N^{b}~,
\end{align}
with $a=2.110(7)$ and $b=-1.009(6)$ for the first transition -eq. \eqref{f1}-, and $a=2.125(4)$ and $b=-0.947(14)$ for the second one -Eq. \eqref{f2}-. 

\section{Conclusions}
\label{secIV}

In this first work of a series of two papers (the second paper is Ref.~\cite{Gamito2022}), we have analyzed the QPT and ESQPTs in the LMG model including in the model Hamiltonian an anharmonic term dependent of the second order Casimir operator of the $u(1)$ system subalgebra, $\left(S+\spin_z\right)\left(S+\spin_z+1\right)$. We have thoroughly studied this system with a focus on its static properties. The second work deals with dynamic aspects and how the ESQPTs modify the system evolution. This work has been fostered by the results obtained considering anharmonicity effects in the 2DVM, and the main motivation to extend this study to the LMG model is twofold. On the one side, the test of different approaches in the LMG model has been pervasive since their definition in nuclear structure studies. It is a simple toy model that offers an excellent playground for approximations and theoretical studies. On the other side, the possibility of accessing experimental realizations of the model provides further interest to studies based on the LMG model.

Apart from defining the aLMG model Hamiltonian and its mean field limit energy functional, we have presented the ground state QPT properties for its energy functional as well as the two ESQPTs that appear in the system for negative values of the anharmonicity parameter $\alpha$. The mean field results for the  ground state QPT properties do not qualitatively change under the inclusion of the anharmonic term in the model Hamiltonian. There are still two phases, a symmetric one and a broken-symmetry phase, with a critical value of the $\xi$ control parameter $\xi_c = 0.2$ that marks a point where a second order ground state QPT takes place. The first ESQPT is linked to the ground state QPT in the sense that it can be explained from the existence of a local maximum in the energy functional once the control parameter goes through its critical value and enters into the system's broken-symmetry phase. However, the second ESQPT is explained from the influence of the anharmonic term on the phase-space boundary of the system, something that in the classical or mean field limit of the system is reflected as a lowering of the asymptotic values of the energy functional for negative values of the $\alpha$ parameter. The influence on the system's spectrum of the two ESQPTs can be seen in the lower panel of Fig.~\ref{ced_figure}, where even and odd parity states are shown: the level piling and different degeneracy patterns depending on the ESQPT phases are clearly evinced, together with the two separatrices. We have deduced from the classical energy functional analytic expressions for the two separatrix lines,  $f_1(\xi,\alpha)$ and $f_2(\xi,\alpha)$, valid on the thermodynamical limit and we have illustrated how there is a threshold value of the $\alpha$ parameter for the second ESQPT to be manifested in the symmetric phase of the system, for $\xi<\xi_c$ (See Fig.~\ref{alpha_thresh_fig}).

We have characterized the aLMG ESQPTs computing the adjacent energy level gap, the expectation value of the $u(1)$ number operator, the participation ratio, and the quantum fidelity susceptibility for various values of the anharmonicity parameter and system sizes. The calculations have been carried out for two values of the  $\xi$ control parameter, $\xi = 0.15$ for a system in  the symmetric phase, and  $\xi = 0.6$ that brings the system into the broken-symmetry phase. The quantities we have used allow to clearly locate both ESQPTs, and the PR shows a strong localization for states with energies close to the critical ESQPT energies. In the ESQPT with separatrix $f_1(\xi,\alpha)$, the localization happens for the $\ket{S,-S}$ basis state, while for the second ESQPT, with separatrix $f_2(\xi,\alpha)$, the localization occurs in the  $\ket{S, S}$ basis state. The QFS is also sensitive to both ESQPTs, having a larger impact on the QFS value the ESQPT occurring at lower energies. 

The existence of ESQPTs has strong implications on system structure and dynamics. Recently, it has been proposed that a conserved quantity can be defined in one of the phases of an ESQPT and the equilibrium values of relevant observables in this phase are dependent on the value of this newly defined constant \cite{Corps2021}. This important result can be applied to the ESQPTs with separatrices $f_1(\xi,\alpha)$ and $f_2(\xi,\alpha)$. Exploring this issue will be a future development of interest with the aLMG.

\begin{acknowledgments}
The authors thank José Enrique García Ramos and Miguel Carvajal Zaera
for fruitful and inspiring discussions on the topic of this paper.
This work is part of the I+D+i projects PID2019-104002GB-C21,
PID2019-104002GB-C22, and PID2020-114687GB-I00 funded by
MCIN/AEI/10.13039/501100011033. This work has also been partially
supported and by the Consejer\'{\i}a de Conocimiento, Investigaci\'on
y Universidad, Junta de Andaluc\'{\i}a and European Regional
Development Fund (ERDF), refs.~UHU-1262561 and US-1380840 and it is
also part of grant Groups FQM-160 and FQM-287 and the project PAIDI
2020 with reference P20\_01247, funded by the Consejería de Economía,
Conocimiento, Empresas y Universidad, Junta de Andalucía (Spain) and
“ERDF—A Way of Making Europe”, by the “European Union” or by the
“European Union NextGenerationEU/PRTR”. Computing resources supporting
this work were provided by the CEAFMC and Universidad de Huelva High
Performance Computer (HPC@UHU) located in the Campus Universitario el
Carmen and funded by FEDER/MINECO project UNHU-15CE-2848.
\end{acknowledgments}

\appendix
\section{The aLMG model density of states}
\label{appa}
As mentioned in the text, when obtaining the density of states for the aLMG model using the Gutzwiller's semiclassical approximation \cite{Gutzwiller2013}, it is handier to recast the aLMG Hamiltonian by replacing the pseudospin $\spin_i$ with a classical angular momentum $j_i$, instead of the $q$ and $p$ generalized coordinate and momentum used in Eq.~\eqref{Hqp}
\begin{equation}
H_{cl}=(1-\xi)\left(j+j_{z}\right)+\frac{\alpha}{2j}\left(j+j_z\right)\left(j+j_z+1\right) +\frac{2\xi}{j}\left(j^2-j_{x}^2\right)~.
\label{eq:LMG2_classical_0}
\end{equation}

We introduce the classical angular momentum azimuthal angle, $\phi = \tan^{-1}(j_y/j_x)\in [0,2\pi)$, and its $z$ component, $j_z\in[-j, j]$, as a valid pair of canonical variables
\begin{equation}
H_{cl}=(1-\xi)\left(j+j_{z}\right)+\frac{\alpha}{2j}\left(j+j_z\right)\left(j+j_z+1\right) +\frac{2\xi}{j}\left(j^2-(j^2-j_{z}^2)\cos^2\phi\right)~.
\label{eq:LMG2_classical_1}
\end{equation}
The density of states Eq.~\eqref{doseq} for the new pair of canonical variables is
\begin{equation}
  \nu(E) = \frac{1}{2\pi}\int_0^{2\pi}d\phi\int_{-j}^{j}  dj_z ~\delta(E-H_{cl}(\phi, j_z)) = \frac{1}{2\pi} \frac{N}{2} \int_0^{2\pi}d\phi\int_{-1}^{1}  dx ~\delta(E-H_{cl}(\phi, x))~,
\end{equation}
where we introduce the rescaled variable $x = j_z/j\in [-1,1]$ and the corresponding Jacobian term, $j = N/2$. Working, as in Eq.~\eqref{enfun}, with the Hamiltonian and energy per particle ${\cal H} = H_{cl}/N$, $\varepsilon = E/N$ and taking into consideration the Dirac delta property $\int dz \delta(z/a) =  |a|\int dz \delta(z)$
\begin{equation}
  \nu(\varepsilon) = \frac{1}{4\pi}\int_0^{2\pi}d\phi\int_{-1}^{1}  dx ~\delta(\varepsilon-{\cal H}(\phi, x))~.
\label{doseq_fin}
\end{equation}

In order to continue, the Dirac Delta composition with a function $f(x)$ is
\begin{equation}
  \delta(f(x))) = \sum_i\frac{\delta(x-x_i)}{\left|\frac{df(x_i)}{dx}\right|} ~,
 \label{deltacomp}
\end{equation}
where the sum extends over all $x_i$, the different roots of $f(x)$. If we apply this property to $f(x) = \varepsilon-{\cal H}(\phi, x)$ in Eq.~\eqref{doseq_fin} we have two possible roots
\begin{equation}
  x_\pm = \frac{\xi-1-\alpha \pm \sqrt{(1-\xi + \alpha)^2-(\alpha + 4 \xi\cos^2\phi) (2 + 2\xi + \alpha - 4\xi\cos^2\phi - 4\varepsilon)}}{\alpha + 4\xi\cos^2\phi}~,
\end{equation}
\noindent though $x_-$ is not a valid solution as it is outside the $x$ variable range. Applying Eq.~\eqref{deltacomp} to  Eq.~\eqref{doseq_fin} for the $x_+$ root we obtain the final result in Eq.~\eqref{analdos}
\begin{equation}
  \nu(\varepsilon) = \frac{1}{4\pi}\bigintss_0^{2\pi}\frac{d\phi}{\left|\sqrt{\left(\frac{1-\xi+\alpha}{2}\right)^2-(\alpha+4\xi\cos^2\phi)\left(\frac{1+\xi}{2}+\frac{\alpha}{4}-\xi\cos^2\phi - \varepsilon\right)}\right|}~.
\label{doseq_almg}
\end{equation}

\bibliography{biblio.bib}
\bibliographystyle{unsrt}

\end{document}